\documentclass[%
 reprint,
superscriptaddress,
 amsmath,amssymb,
 aps,
 prl
]{revtex4-2}

\usepackage{graphicx}% Include figure files
\usepackage{dcolumn}% Align table columns on decimal point
\usepackage{bm}% bold math
\usepackage{tabularx}
\usepackage{blindtext}
\usepackage[colorlinks]{hyperref}
\hypersetup{
 colorlinks=true,
 citecolor=red,
 linkcolor=blue,
 urlcolor=cyan}

\usepackage{amsmath}
\usepackage[flushleft]{threeparttable}

\usepackage{xparse}
\usepackage{subcaption}
\NewDocumentCommand{\tens}{t_}
 {%
  \IfBooleanTF{#1}
   {\tensop}
   {\otimes}%
 }
\NewDocumentCommand{\tensop}{m}
 {%
  \mathbin{\mathop{\otimes}\displaylimits_{#1}}%
 }

\usepackage{titlesec}
\titlespacing*{\section}
{0pt}{1ex plus 1ex minus .2ex}{1ex plus .2ex}
\titlespacing*{\subsection}
{0pt}{1ex plus 1ex minus .2ex}{1ex plus .2ex}
\usepackage{amsmath, amssymb, graphicx, float, amsfonts, bm, mathrsfs}
\usepackage{float}

\renewcommand{\ao}{ \hat{a} }
\newcommand{\ad}{ \hat{a}^\dag }
\newcommand{\bo}{ \hat{b} }
\newcommand{\bd}{ \hat{b}^\dag }
\newcommand{\co}{ \hat{c} }
\newcommand{\cd}{ \hat{c}^\dag }
\newcommand{\Bo}{ \hat{B} }
\newcommand{\Bd}{ \hat{B}^\dag }
\newcommand{\w}{ \omega }

\newcommand{\Bin}{ \hat{B}^\textrm{in} }
\newcommand{\Bind}{ \hat{B}^{\textrm{in} \dag} }
\newcommand{\Bout}{ \hat{B}^\textrm{out} }
\newcommand{\Boutd}{ \hat{B}^{\textrm{out} \dag} }
\newcommand{\Cin}{ \hat{C}^\textrm{in} }

\newcommand{\Cout}{ \hat{C}^\textrm{out} }

\newcommand{\bK}{ \bm{\mathcal{K}} }
\newcommand{\bH}{ \bm{\mathcal{H}} }

\newcommand{\sC}{ \mathcal{C} }

\newcommand{\sE}{ \mathcal{E} }
\usepackage{soul}
\usepackage{ulem}
\usepackage{xcolor}

\begin{document}

\preprint{APS/123-QED}

\title{Efficient in-situ generation of photon-memory entanglement in a nonlinear cavity}% Force line breaks with \\

%  \email{Second.Author@institution.edu}
% \affiliation{%
%  Authors' institution and/or address\\
%  This line break forced with \textbackslash\textbackslash
% }%

% \collaboration{MUSO Collaboration}%\noaffiliation

\author{Hoi-Kwan Lau}
\affiliation{Pritzker School of Molecular Engineering, University of Chicago, Chicago, Illinois 60637, USA}%
\affiliation{Department of Physics, Simon Fraser University, Burnaby, BC V5A 1S6, Canada}%

\author{Hong Qiao}
\affiliation{Pritzker School of Molecular Engineering, University of Chicago, Chicago, Illinois 60637, USA}%

\author{Aashish A. Clerk}
\affiliation{Pritzker School of Molecular Engineering, University of Chicago, Chicago, Illinois 60637, USA}%

\author{Tian Zhong}
\affiliation{Pritzker School of Molecular Engineering, University of Chicago, Chicago, Illinois 60637, USA}%
\email{tzh@uchicago.edu}

\date{\today}

%              %  but any date may be explicitly specified

\begin{abstract}
Parametrically driving an optical cavity that simultaneously couples to an atomic ensemble quantum memory enables in-situ generation of multimode photon-memory entanglement. A high-rate bi-party photon-memory entanglement can be generated even after discarding one entangled optical mode. This protocol can be realized with existing technologies based on photonic resonators integrated with a rare-earth-ion doped quantum memory. The proposed scheme shows significant advantages in entanglement generation rates compared with prevailing quantum memory protocols and experiments, with theoretical Ebit rates of tens of MHz without fine-tuned operating conditions. Such a photon-memory entanglement source offers a versatile resource for quantum networking and interconnect applications.
\end{abstract}

\maketitle

\textit{Introduction.-}
Entanglement between itinerant optical photons and matter degrees of freedom is a quintessential ingredient for remote quantum interconnects, with imminent applications in quantum networks \cite{kimble_quantum_2008,wehner_quantum_2018, Cirac1997}, quantum transduction \cite{MOconvert,mirhosseini_quantum_2020}, quantum-enhanced telescopes \cite{gottesman2012longer}, and distributed quantum computation \cite{grover1997quantum, Jiang2007, Monroe2014}. Using Raman transitions, the well-known Duan-Lukin-Cirac-Zoller (DLCZ) protocol generates one entangled bit (Ebit) between two remote atomic ensembles, with built-in entanglement purification and noise resilience \cite{duan_long-distance_2001}. This protocol was extended to multimode operation by combining an EPR photon pair source with a broadband quantum memory realized via inhomogeneously broadened atomic ensembles in solids, leading to orders of magnitude speedup of the Ebit generation rate and reduced experimental complexities \cite{simon_quantum_2007}. While these protocols are long established, experimental demonstrations of multi-mode entanglement between two remote atomic memories were only achieved very recently \cite{lago-rivera_telecom-heralded_2021, liu_heralded_2021}. Long-standing challenges hampering this scheme include photon loss and bandwidth mismatch at the interface between itinerant optical modes and the stationary memory degree of freedom. For instance, in the experiments employing separate photon sources and memories, the entanglement generation rates are limited even over a short fiber distance \cite{lago-rivera_telecom-heralded_2021, liu_heralded_2021}. One strategy to mitigate this loss at the light-matter interface is to use an atom-containing, high-finesse optical cavity with a fine-tuned impedance matching (i.e. unit coupling coopertivity) \cite{afzelius_impedance-matched_2010, moiseev_efficient_2010}. However, such cavity-enhanced memory interfaces \cite{Sabooni2013, Jobez2014, Akhmedzhanov2016, Minnegaliev2021} would require a matching narrow-band photon source, which not only adds significant system complexities, but also constraints the overall entanglement generation bandwidth and throughput.

In this Letter we propose a scheme for efficient multimode entanglement generation between an optical photon and an atomic memory within a nonlinear cavity. By parametrically driving the cavity, we produce  non-degenerate photon pairs (i.e.~signal and idler photons), with one of the photons (i.e.~signal) having spectral overlap with an inhomogeneously broadened atomic ensemble. 
By constructing an equivalent circuit representation of the steady-state input-output relations, we uncover the basic structure of the resulting multi-partite entanglement generated between the atomic and two photonic degrees of freedom.
We find that our scheme is capable of generating photon-memory entanglement over a broad spectrum of modes, some of which can involve highly excited states. We quantify the performance of our system by the entanglement rate, which is exact beyond the weak excitation regime and upper-bounds the performance of any practical entanglement distribution protocol. Our scheme demonstrates robust and efficient entanglement creation without requiring cavity impedance matching, parameter fine tuning or strong coupling. 
Even with an imperfect intra-cavity photon-memory transfer and discarding the signal photons, we see no significant degradation of bipartite entanglement between the memory and the idler photons. 

We further describe experimental realizations using existing technologies based on photonic microcavities coupled to ensembles of rare-earth ions doped in a crystalline substrate. The proposed system is capable of generating up to 50 MHz Ebits over a range of accessible experimental parameters. The high Ebit generation rate is result of a large number of frequency modes within the inhomogeneous linewidth of the memory, and the accurate accounting of entanglement in the strongly driven regime.
Our proposal is distinct from the DLCZ protocol with its quasi-CW, multimode operation, while it eliminates the loss and mismatch challenges in pair-source based repeater protocols \cite{simon_quantum_2007}. Furthermore, the intrinsic tri-partite nature of the entanglement - between two photonic and one memory degrees of freedom - opens future opportunities of generating and distributing multi-partite entanglement over a network for quantum secret sharing \cite{2002PhRvA..65d2310T, Lau:2013wb}, multi-partite teleportation \cite{lian2007continuous}, and distributed sensing \cite{zhuang2018distributed}. 

\begin{figure}[t]
  \includegraphics[width=1\linewidth]{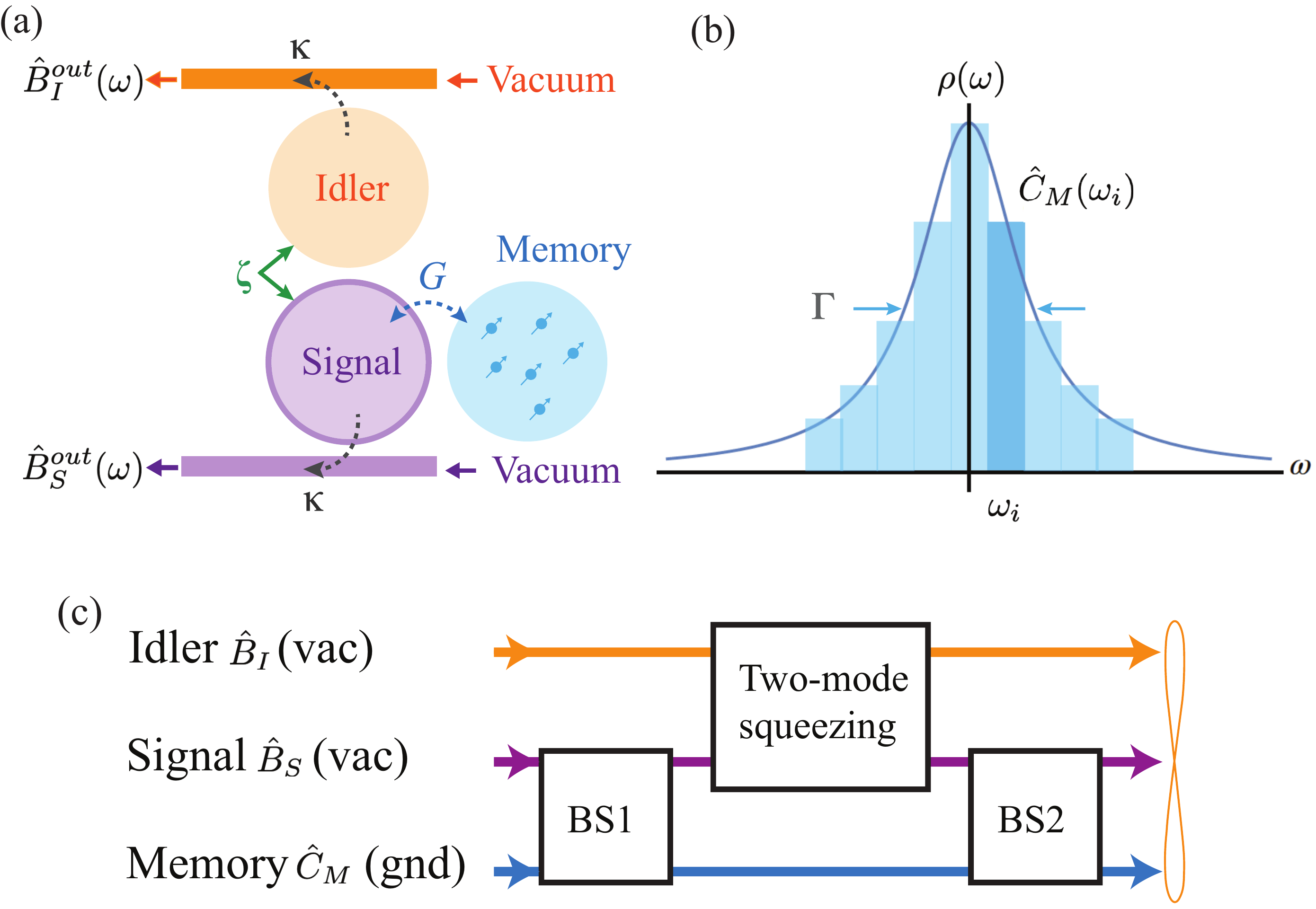}
\setlength{\belowcaptionskip}{-8pt}
\caption{(a) Generalized model of three-mode system: Entangled signal ($\hat{B}_S$) and idler ($\hat{B}_I$) are generated by parametric driving and signal is coupled to atomic memory ($\hat{C}_M$) (b) Spectral slicing of inhomogeneously broadened atomic ensemble memory. (c) Equivalent circuit consists of one two-mode-squeezing gate sandwiched by two beam-splitters.  The amount of output excitation is characterized by the TMS strength $\cosh^2 r$. The second beam-spliter (BS2), which determines the amount of entanglement or photon going into memory, has a beam-splitting angle $\theta_2$.  The first beam splitter (BS1), which has a beam-splitting angle $\theta_1$, is irrelevant for our situation of vacuum/ground state input.}
\label{cavity}
\end{figure}

% modeling of the system
\textit{Entanglement generation.}- Our entanglement generation process can be modelled as two cavity modes, signal ($\hat{B}_S$) and idler ($\hat{B}_I$), which are spectrally distinct, and which couple independently to a common waveguide with a coupling rate $\kappa$. The waveguide frequency mode, that is detuned by $\w$ from the signal (idler) mode frequency, is represented by $\bo_{S\w}$ ($\bo_{I\w}$). The signal mode also resonantly couples to an atomic ensemble memory %($\hat{C}_{M}$) 
with a spectral distribution $\rho(\w)$ (i.e. the signal cavity frequency coincides with the center of the atomic ensemble inhomogeneous distribution).  In the rotating frame of the cavity modes the Hamiltonian of the system is
\begin{eqnarray}
\hat{H} &=& \xi (\hat{B}^{\dagger}_S \hat{B}^{\dagger}_I + \hat{B}_S \hat{B}_I) + \int \w \cd_{M\w} \co_{M\w} d\w   \nonumber \\
&& + G \int \sqrt{\rho(\w)} \left(\hat{B}^{\dagger}_S \co_{M\w} +  \cd_{M\w} \hat{B}_S \right) d\w~,
\end{eqnarray}
\noindent where $G$ is the collective atom-photon coupling. In our protocol the device is driven by a constant intensity pump at the frequency $\nu_p = (\nu_S + \nu_I)/2$ from $t=t_o$ to $t=t_f$, this will induce a parametric drive with strength $\xi$ in the cavities. Generally, atoms with the frequencies within the `strip' $\omega$ to $\omega + d\omega$ are not resolvable if $d\omega$ is much smaller than the resolvable bandwidth $\delta_r \equiv 1/(t_f - t_o)$. We consider the ensemble has a wide inhomogeneous broadening $\Gamma \gg \delta_r$, so the collective atomic excitation within each strip can be represented by the bosonic operator $\cd_{M\w}$, and the collective atom-photon coupling of each strip is scaled by $\sqrt{\rho(\omega)}$ \cite{seeSI}. We note that this representation is compatible with holeburning memory protocols such as atomic frequency combs \cite{afzelius_multimode_2009}, in which the comb spacing $\Delta$ cannot be resolved during the pump time, i.e. $\delta_r \gg \Delta$, and the ensemble can be effectively modelled with a modified atomic density.

For sufficiently long pump time, i.e. $\tau_{\textrm{pump}} \equiv t_f - t_o \gg 1/\kappa$, the output photonic modes and atoms can be characterized by the steady-state scattering relation
\begin{eqnarray}\label{eq:3io2}
\Boutd_I[\w] &=& T_{I I} \Bind_I[\w] + T_{I S} \Bin_S[\w] + T_{I M} \Cin_M[\w] \nonumber \\
\Bout_S[\w] &=& T_{S I} \Bind_I[\w] + T_{S S} \Bin_S[\w] + T_{S M} \Cin_M[\w] \\
\Cout_M[\w] &=& T_{M I} \Bind_I[\w] + T_{M S} \Bin_S[\w] + T_{M M} \Cin_M[\w] ~. \nonumber
\end{eqnarray}
We note that the scattering amplitudes $T$ depend on detuning $\omega$ \cite{seeSI}, which is omitted in the above equations only for brevity.
The input/output operators of the photonic waveguide follow the standard definitions, $\hat{B}^{\textrm{in}/\textrm{out}}_u(t)\equiv \frac{1}{\sqrt{2\pi}}\int \bo_{u\w}(t_o/t_f) e^{-i \w (t-t_o/t_f)}d\w$, for $u=S, I$.  Analogously, we can define the effective input and output operators of atoms as $\hat{C}^{\textrm{in}/\textrm{out}}_M(t)\equiv \frac{1}{\sqrt{2\pi}}\int \co_{M\w}(t_o/t_f) e^{-i \w (t-t_o/t_f)}d\w$, which characterizes the transformation of the atomic state before and after interacting with the cavity.  

This scattering relation shows that our system intrinsically generates tri-partite entanglement between each frequency mode of idler, signal, and memory. The equivalent circuit is depicted in Fig.~\ref{cavity}(c), which involves two beam splitters (BS) with the same angle $\tan\theta_1 = \left|\frac{T_{I M}(\w)}{T_{I S}(\w)} \right|=\tan\theta_2 = \left| \frac{T_{M I}(\w)}{T_{S I}(\w)} \right|$, and a two-mode-squeezing (TMS) interaction with strength $\cosh^2 r = |T_{I I}(\w)|^2$. We note that the mode transformation denoted by Eq.~(\ref{eq:3io2}) applies to any input state, so our setup can also be applied to engineering logic gates between itinerant photons and quantum states stored in the memory \cite{Lau:2019bs}. For our current goal of in-situ generation of photon-memory entanglement, we assume both signal and idler inputs are vacuum and the memory is initialized in the ground state.

\begin{figure}[t]
  \includegraphics[width=1\linewidth]{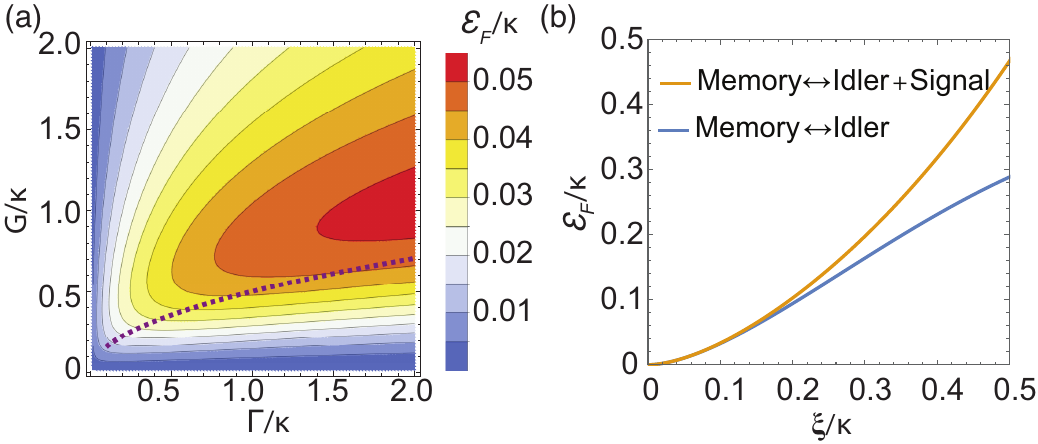}
\setlength{\belowcaptionskip}{-8pt}
\caption{(a) Normalized entanglement rate $\sE_F/\kappa$ as a function of the coupling $G$ and the atomic ensemble linewidth $\Gamma$ while keeping a weak parametric drive at $\xi =0.1 \kappa$. The cavity impedance matching condition is also plotted as in red curve. (b) Normalized entanglement rate $\sE_F/\kappa$ as a function of parametric drive $\xi/\kappa$. Here we assumed 2$G$=$\Gamma$=$\kappa$, so $\sC=1$. Orange (Blue) curve represents the entanglement between memory and idler output only (both signal and idler output).}
\label{fig:ER}
\end{figure}

To quantify the amount of entanglement, we consider a Lorentzian spectrum, $\rho(\w) = \frac{1}{2\pi} \frac{\Gamma}{\w^2 + (\Gamma/2)^2}$, with an ensemble inhomogeneous broadening $\Gamma$.  
Based on the equivalent circuit in Fig.~\ref{cavity}(c), one might guess that a complete transfer of the entangled signal photon to the memory necessarily requires a large effective beam splitting angle
\begin{equation}\label{eq:theta2}
\tan^2\theta_2 = \left| \frac{T_{M I}(\omega)}{T_{S I}(\omega)}\right|^2 =  \frac{\sC}{1 + 4\w^2/\Gamma^2}.
\end{equation}
This would require in turn a large collective atomic ensemble-cavity cooperativity $\sC \equiv 4 G^2/\kappa \Gamma \gg 1$. We find however that this condition is neither necessary nor optimal for our memory.  Heuristically, increasing $\sC$ enhances the effective damping of the cavity, which is harmful as it suppresses photon generation due to the parametric drive.

Further, our analysis indicates that there is no special utility in achieving impedance matching (i.e. $\sC=1$):  we thus conclude that fine tuning to reach this matching is not required. To be more specific, we quantify the performance of our protocol by the entanglement rate $\sE_R$, which is the total amount of photon-memory entanglement generated in all frequency modes per unit time \cite{2016NJPh...18f3022J, Zhong:2020fp}:
\begin{equation}\label{eq:ER}
\sE_R \equiv \frac{\sum_{\w} \sE_F(\w) }{\tau_{\textrm{pump}}}\approx \frac{1}{2\pi} \int \sE_F(\w) d\w~.
\end{equation}
We note that the entanglement between modes with different frequencies is negligible as long as $\tau_{\textrm{pump}} \gg 1/\kappa, 1/\Gamma$ \cite{zhuang2018distributed}. We account for the entanglement of each frequency mode at $\w$ by entanglement of formation $\sE_F(\w)$ because it has clear operational meaning and can be computed exactly for our output state that is Gaussian and balanced.  Our choice of $\sE_R$ as entanglement quantifier is advantageous as it does more than account for the entanglement associated with single excitations:  contributions from vacuum and highly excited states are also fully included. Furthermore, it represents the device's maximum performance in entanglement generation for a fixed choice of system parameters, i.e. any post-selection or non-Gaussian process cannot improve the rate of entanglement generation.

%and $\sE_F(\w)$ is the entanglement of formation of the frequency mode at $\w$, which can be computed exactly as the output state is Gaussian and balanced \cite{2017PhRvA..96f2338T}. 

Fig.~\ref{fig:ER}(a) plots the memory-idler entanglement rate $\sE_R$ as a function of both the  atomic ensemble linewidth and the ensemble-cavity coupling.
%$\Gamma$, with a perfect impedance matching condition indicated as the red curve. 
For maximizing $\sE_R$, it is clear that impedance matching is neither necessary nor universally optimal. We also observe that around the regime of maximum performance, the value of $\sE_R$ is robust against order-unity changes in system parameters. This is ultimately because the memory captures the entangled photon directly from the signal cavity, a process that can occur efficiently without the fine tuning that would be required if instead photons had to be captured from a waveguide.  

The above analysis considered memory-idler photon entanglement assuming that outgoing signal photons are discarded. One might worry that this discarding results in a significant loss of entanglement, especially since the effective beam splitting angle in Eq.~(\ref{eq:theta2}) is not overwhelmingly large at the parameters of maximum performance in Fig.~\ref{fig:ER}(a) (i.e.~$\sC$ is of order unity). To investigate this concern, we compare in Fig.~\ref{fig:ER}(b) the entanglement rate for two cases: 
(i) between memory and both photonic (signal and idler) outputs (computed using the entropy of the final memory state); (ii) between memory and idler output only, with the signal output discarded.
Interestingly, the atom-photon entanglement rate is barely degraded when the drive is weak (i.e. $\xi \ll \kappa$), and shows only about a factor of two reduction when $\xi$ is comparable to $\kappa$ (instability occurs when $\xi \rightarrow \kappa/\sqrt{2}$ in our choice of parameters \cite{seeSI}). The intuition behind this effect is that the entanglement rate has significant contributions from off-resonant modes (i.e. $\omega \neq 0$). These modes are only weakly excited even when the resonant mode is driven close to instability. On the other hand, idler-memory entanglement would only be destroyed when a photon is detected in signal output. Since the signal output of off-resonant modes contain mainly vacuum, there is a high probability that the idler-memory entanglement is retained even if signal is monitored or discarded. Nevertheless, despite the weak excitation, the tri-partite entanglement between idler, signal and memory is genuine and can be harnessed as a resource for quantum information applications \cite{2002PhRvA..65d2310T, Lau:2013wb, lian2007continuous, zhuang2018distributed}.

\begin{figure}[t]
  \includegraphics[width=1\linewidth]{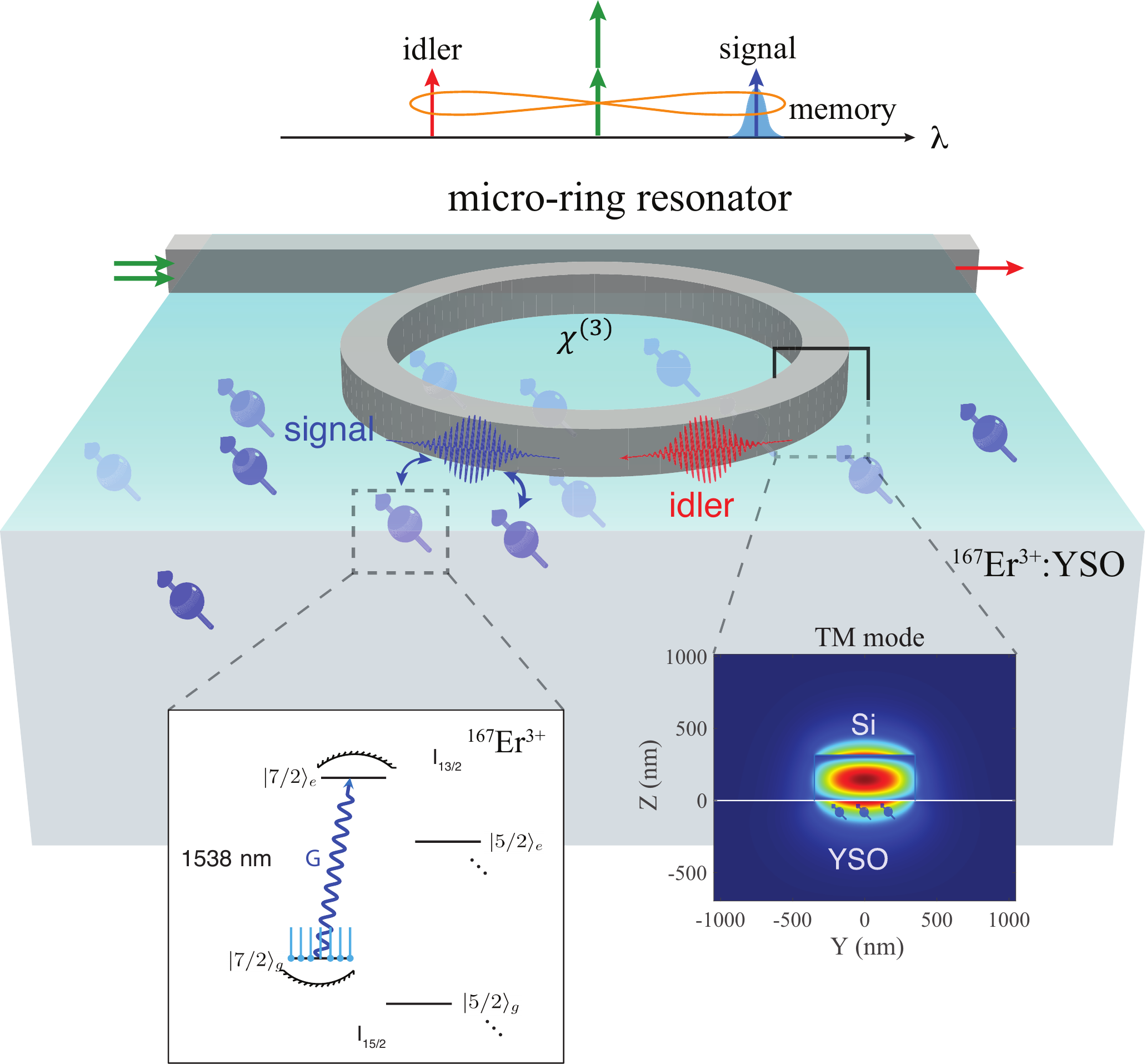}
\setlength{\belowcaptionskip}{-8pt}
\caption{Experimental realization of in-situ entanglement generation based on a micro-ring resonator on Er:YSO crystal. The $\rm {}^{167} Er^{3+}$ ions are evanescently coupled to optical field confined in a $\rm Si$ waveguide. A pump field (green) generates entangled signal (blue) and idler (red) photons pairs via sFWM process while signal photons are coupled with Er optical transitions and the idler photons are coupled out of the cavity to the waveguide. Left inset shows the quantum memory scheme based on AFC and the relevant $\rm {}^{167} Er^{3+}$ energy levels. Right inset shows the simulation of transverse magnetic mode profile of $\rm Si$ waveguide on a $\rm {}^{167} Er^{3+}:YSO$ substrate.}
\label{Experiment}
\end{figure}

\textit{Experimental scheme.}- The proposed in-situ entanglement generation can be realized experimentally with existing technologies. We design an integrated device that consists of a silicon micro-ring resonator on top of a $\rm {}^{167} Er^{3+}$:$\rm Y_2SiO_5$ crystalline substrate, as shown in Fig.~\ref{Experiment}. Entangled signal and idler photon pairs are generated by spontaneous four-wave-mixing (sFWM) in the silicon resonator with an estimated MHz generation rate at $\mu$W pump power \cite{azzini_ultra-low_2012,jiang_silicon-chip_2015} \cite{seeSI}. The on-chip micron-scale resonator implementation is necessary for a sufficiently large (i.e.~$\geq$ 1 nm) free spectral range \cite{seeSI} so that the signal, idler and pump photons can be efficiently separated with dense wavelength-division-multiplexing (DWDM) filters. The signal cavity mode is resonant with the $\rm {}^{167} Er^{3+}:YSO$ optical transition at 1539 nm, between the ground and excited-state $|m_s=7/2\rangle_{e-g}$ hyperfine levels \cite{rancic_coherence_2018}. Assuming a doping concentration of 17 parts per million (ppm), the evanescent coupling between the transverse magnetic (TM) mode of the resonator and the Er ensembles leads to a collective coupling strength of $G=2\pi\times0.173$ GHz \cite{seeSI}.

The quantum memory protocol we adopt is based on optical atomic frequency combs (AFC), which have built-in multimode capability \cite{afzelius_multimode_2009}. A long-lived AFC is first prepared by initializing and holeburning on the $\rm {}^{167} Er^{3+}$ $|m_s=7/2\rangle_{e-g}$ transition prior to entanglement generation. Then a quasi-CW pump laser is turned on with a time duration $\tau_{\textrm{pump}} \gg 1/\kappa$ to continuously generate entangled photons in idler and signal modes. If there is no memory, the output is a continuum of entangled photon with frequency centered at $\nu_I$ and $\nu_S$ respectively, which can also be viewed as a continuous stream of temporally (or time-energy) entangled photons with the duration of each defined by the cavity lifetime $1/\kappa$. In our scheme that involves additionally the memory, photons in the signal cavity mode are stored into the AFC with a storage time $T_M=1/\Delta$ where $\Delta$ is the AFC comb spacing. We further require that the pump duration is shorter than the storage time, so that our ensemble description of the memory remains valid.

AFC memory storage efficiency in our scheme is determined by Eq.~(\ref{eq:theta2}), in which the effective cooperativity should take into account the finesse of the AFC: $F=\Delta/\gamma$, where $\gamma$ is the spectral width of the comb tooth. For $\rm {}^{167}$Er:YSO, a storage time of $1\mu s$ is achievable, with a AFC finesse $F$=3, $\Delta$=1 MHz, and $\gamma\approx2\pi\times$0.3 MHz \cite{Yasui21, Craiciu2019}. Long-term storage is also feasible with spin-wave memory \cite{PhysRevLett.87.173601} using an additional hyperfine level \cite{rancic_coherence_2018}. While the signal photons are stored, the idler photons are propagating out of the cavity. The average idler photon number in each temporal mode is controlled by adjusting the pump strength $\xi$. We stress that the entanglement generated in our scheme can be accounted for even when $\xi$ is large.

\begin{table}[t]
\small
\centering
  \caption{Example experimental parameters}
  \begin{tabularx}{0.5\textwidth}{p{1.5cm}p{1.5cm}p{1.5cm}p{1.5cm}p{1cm}p{1cm}}
    \hline\hline
    $\Gamma$/2$\pi$  & $G^{\rm AFC}$/2$\pi$ & $\kappa$/2$\pi$  & $\xi$/2$\pi$ & $\rm F$ & $\rm T_{M}$\\
    (MHz) & (MHz) & (MHz) & (MHz) & & ($\mu$s)\\
     150  & 100 & 100 & 80 & 3 & 1 \\
    \hline\hline
  \end{tabularx}
  \label{tab1}
\end{table}

\begin{figure}[t]
  \includegraphics[width=\linewidth]{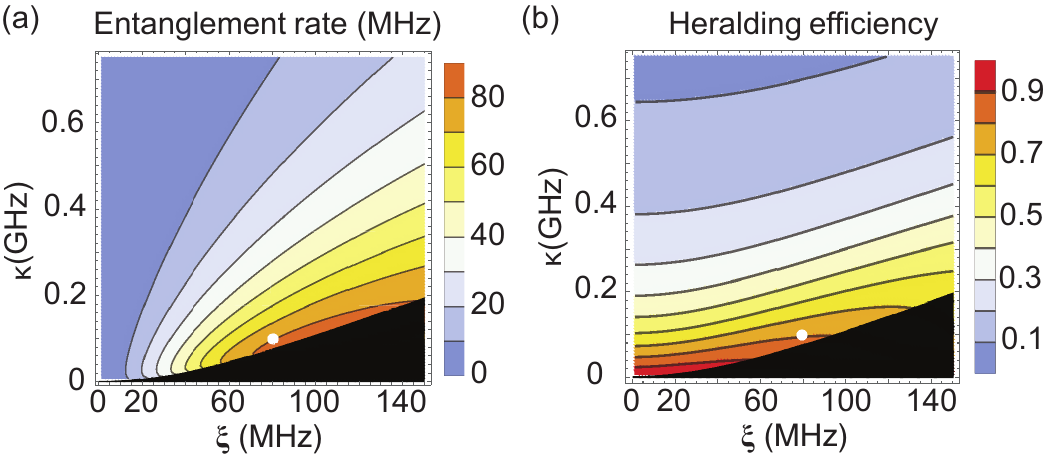}
\caption{(a) Continuous-variable (CV) entanglement rate $\sE_R$ of generated photon-memory pair and (b) Heralding efficiency $\eta$ of memory excitation per detected idler photon for discrete-variable (DV) entanglement generation. The white dots indicate system parameters given in Table~\ref{tab1}.}
\label{fig:HE}
\end{figure}

\textit{Entanglement rate.} - We first analyze the entanglement rate of our experimental setup. Remote entanglement generation with separate photon pair source and quantum memory has demonstrated heralded entanglement rate up to 15.6 kHz \cite{lago-rivera_telecom-heralded_2021}, which is mainly limited by a low photon pair generation rate, high loss and signal photon storage inefficiency. Our in-situ scheme can circumvent the limitations of interface loss and storage inefficiency through directly coupling to memory. Furthermore, the entanglement rate $\sE_R$ takes into account the entanglement of highly excited modes, so the performance of our setup can be faithfully quantified even the cavity is strongly driven. 

In Fig.~\ref{fig:HE}(a) with a fixed coupling $G^{\rm AFC}=G/\sqrt{F}=2\pi\times0.1$ GHz, we show the entanglement rate at varying parametric drive $\xi$ and cavity decay rate $\kappa$. The peak entanglement rate can reach as much as 85 MHz Ebits. For the example experimental parameters in Table.~\ref{tab1}, the entanglement rate at $\xi$=2$\pi\times$80 MHz (white dot) is 79 MHz. This high Ebit rate is achieved when the cavity is driven just below the instability threshold, that is the black region in Fig.~4. 
Although the resonant frequency mode will be highly excited near the threshold, the time-bin mode that is addressed by usual time-resolved detectors contains much less than one photon on average, due to the contribution of the off-resonant frequency modes which are weakly excited. This combination of in-situ generation and accounting the full structure of entanglement shows that our setup has the potential to achieve orders of magnitude higher rates than that of latest experiments in \cite{lago-rivera_telecom-heralded_2021, liu_heralded_2021}.

\textit{Heralding efficiency.}- The inherently continuous-variable entanglement in our scheme can also be a resource for discrete-variable encoding. One use case is heralded entanglement generation in a quantum repeater network, in which detection of an itinerant photon heralds the storage of a qubit in the memory. We here analyze the heralding efficiency as an important figure of merit in such context. Previously, Ref.~\cite{ramelow_highly_2013} demonstrated heralded single-photon source with 83\% heralding efficiency. However, this efficiency would be much lower if one needs to store the photon into a quantum memory (19\% demonstrated in \cite{lago-rivera_telecom-heralded_2021} and 14.3\% AFC efficiency in a two-photon-detection heralded entanglement distribution experiment \cite{liu_heralded_2021}). We thus define the heralding efficiency as rate of non-zero memory excitation to the output rate of non-zero idler photon, 
\begin{equation}
    \eta \equiv \frac{\int \left(1-\frac{1}{1+ |T_{MI}(\omega)|^2}\right) d\w}
    {\int \left(1-\frac{1}{1+|T_{IS}(\omega)|^2 + |T_{IM}(\omega)|^2}\right) d\w}~.
\end{equation}
We plot the heralding efficiency as a function of $\kappa$ and $\rm \xi$ in Fig.~\ref{fig:HE}(b). With a $G^{\rm AFC}=2\pi\times0.1$ GHz, $\Gamma=2\pi\times150$ MHz, $\xi=2\pi\times10$ MHz, a heralding efficiency up to 99\% can be achieved with a $\kappa=2\pi\times3$ MHz. With more practical parameters in Table I, we obtain a heralding efficiency of 79\%.

\textit{Memory retrieval.}- The stored entanglement eventually needs to be retrieved. Here we outline, without loss of generality, two retrieval strategies. First, the entanglement in the memory can be retrieved by transducing to and readout through an auxiliary non-optical mode. For instance, coupling the atomic ensemble to a proximal superconducting resonator allows transduction of the memory excitation to microwave photons following the protocol in \cite{OBrien2014}. The entanglement retrieval efficiency in this case is determined by the transduction efficiency, which can theoretically reach unity \cite{williamson_magneto-optic_2014}. The second strategy retrieves the entanglement through an optical mode - for convenience we consider through the same cavity mode in which entanglement was originally generated. A perfect retrieval is only achieved when all the memory spectral modes fulfils the impedance matching condition. While a standard impedance matched cavity only guarantees perfect retrieval for the memory frequency modes within a fraction ($\approx1/3$) of the cavity linewidth \cite{moiseev_efficient_2010, Moiseev_2021}, \cite{Moiseev_2021} proposed using dispersion compensation to enlarge this impedance matching bandwidth beyond the cavity $\kappa$ towards the full ensemble inhomogeneous linewidth. In our scheme, we can realize a similar extended cavity impedance matching by dynamically switching the cavity to impedance matching point (e.g. via carrier injection to switch the Q) and rapidly sweeping the cavity resonance frequency across the ensemble inhomogeneous broadening. The integrated nanophotonic platform is amenable for such active cavity switching and fast chirping, which has been realized by incorporating high-speed field modulation (up to 10 GHz in LiNbO$_3$ resonators \cite{Wang:18}) or current injection \cite{Xu2005} elements to the resonator. Finally, even without a perfect fidelity, the entanglement extracted from the memory, when optimized for a high throughput generation, is still a desirable resource for quantum network as imperfect entanglement would be distilled during the entanglement purification step \cite{PhysRevA.59.169, deutsch1996quantum, bennett1996purification} of  quantum repeater protocols. Notably, using the same parameters in Table I and assuming no memory errors, we find that 81\% of the generated entanglement can be retrieved \cite{seeSI}, and this efficiency is further optimized when operating at the exact impedance matching condition at an expense of entanglement generation rate.

%\textit{Summary.-} We have presented a scheme for in-situ generation of entanglement between itinerant optical photons and quantum memories and describe an experimental realization based on existing technologies. By eliminating loss associated with the photon-memory interfaces while retaining the multimode feature of the ensemble-based memories, we demonstrate a system capable of generating robust, 10s MHz-rate, bi-party entanglement between optical photons and atomic memories without impedance matching nor fine parameter tuning.  The general framework of our proposal can be extended to other hybrid quantum systems. With ensembles of spins in solids, coupling to superconducting resonators would enable high-rate microwave-memory entanglement. Similar idea is applicable to optomechanics to generate photon-phonon entanglement \cite{wang_bipartite_2015,wallucks_quantum_2020,zhong_proposal_2020}, and to magneto-optical systems for photon-magnon entanglement \cite{tanji_heralded_2009,simon_single-photon_2007, williamson_magneto-optic_2014}. Finally, the intrinsic three-mode entangling operation in our scheme also opens a future possibility to distribute hybrid multi-party entanglement across a network for advanced quantum communication and sensing protocols.

We acknowledge helpful discussions with Kurt Jacob. This work was funded by a National Science Foundation (NSF) Faculty Early Career Development Program (CAREER) Grant (No.~1944715), partially supported by the University of Chicago Materials Research Science and Engineering Center, which is funded by the National Science Foundation under award number DMR-2011854, and the NSF QLCI for Hybrid Quantum Architectures and Networks (NSF Grant No. 2016136). H.-K.L. acknowledges support from Canada Research Chairs (CRC- 2020-00134).

\bibliography{arXivV2}  {}{}
\bibliographystyle{apsrev4-2}

\clearpage

\onecolumngrid

\begin{center}
    \mbox{\Large \textbf{Supplemental Material}}
\end{center}

\subsection{Derivation of scattering relations}

We consider a system consisting of two cavity modes, signal ($\hat{B}_S$) and idler ($\hat{B}_I$), of which their resonant frequencies are respectively $\nu_S$ and $\nu_I$.  The cavity is driven by a pump with frequency $\nu_p = (\nu_S+\nu_I)/2$, which induces a parametric interaction on the modes due to cavity's intrinsic non-linearity.  The cavity couples to a waveguide.  Since our operation time is much longer than the round-trip time of the optical cavity, i.e. $t_f-t_o \gg 1/|\nu_I - \nu_S|$ , each cavity mode couples mainly to the waveguide frequency modes around its resonant frequency, and thus the waveguide can be modelled as two continuum of modes centered at $\nu_S$ and $\nu_I$.  The signal mode is further coupled to an atomic ensemble memory that consists of $N$ atoms.  In the rotating frame with respect to the cavity mode frequencies, the total Hamiltonian is given by
\begin{eqnarray}\label{eq:Full_H}
%H &=& \omega_s \Bd_S \Bo_S + \w_I \Bd_I \Bo_I + \xi (\Bd_S \Bd_I + \Bo_S \Bo_I) +\sum_n \w_n \sigma^{(n)}_+ \sigma^{(n)}_- + \int \w_s \bd_{\w_s}\bo_{\w_s} d\w_s + \int \w_I \bd_{\w_I}\bo_{\w_I} d\w_I \nonumber \\
%&&+ \sqrt{\frac{\kappa_s}{2\pi}} \int (\ad_s \bo_{\w_s} +  \bd_{\w_s}\ao_s) d\w_s + \sqrt{\frac{\kappa_I}{2\pi}} \int (\ad_I \bo_{\w_I} +  \bd_{\w_I}\ao_I) d\w_I + g \sum_n (\ad_s \sigma^{(n)}_- +  \sigma^{(n)}_+ \ao_s)~.
\hat{H}_\textrm{full} &=& \xi (\Bd_S \Bd_I + \Bo_S \Bo_I)  + \int \w \bd_{S\w}\bo_{S\w} d\w + \int \w \bd_{I\w}\bo_{I\w} d\w +\sum_n \w_n \hat{\sigma}^{(n)}_+ \hat{\sigma}^{(n)}_- \nonumber \\
&&+ \sqrt{\frac{\kappa_S}{2\pi}} \int (\Bd_S \bo_{S\w} +  \bd_{S\w}\Bo_S) d\w + \sqrt{\frac{\kappa_I}{2\pi}} \int (\Bd_I \bo_{I\w} +  \bd_{I\w}\Bo_I) d\w + g \sum_n (\Bd_S \hat{\sigma}^{(n)}_- +  \hat{\sigma}^{(n)}_+ \Bo_S)~.
\end{eqnarray}
where $\xi$ is the strength of the parametric coupling; 
$\bo_{S\w}$ and $\bo_{I\w}$ are the annihilation operators of the waveguide frequency modes that are $\w$ detuned from the signal and idler cavity modes, respectively; 
$\omega_n \equiv \nu_n - \nu_S$ is the detuning between the energy level splitting $\nu_n$ of the $n^{th}$ atom and the signal mode frequency; $\hat{\sigma}^{(n)}_-$ and $\hat{\sigma}^{(n)}_+$ are respectively the lowering and raising operators of the $n$th atom; 
$\kappa_S$ and $\kappa_I$ are respectively the waveguide coupling rates of the signal and idler cavity modes;
the coupling strength between the atom and signal mode is assumed to be homogeneous and is given by $g$. Because the cavity-waveguide coupling rate is assumed to be much weaker than the cavity mode frequency difference, the off-resonant waveguide modes are barely excited and so the integral domains of the detuning $\w$ can be well approximated as from $-\infty$ to $\infty$. For simplicity, we have neglected all internal losses, but they can be straightforwardly incorporated and will be analyzed in future works.

We assume the spectrum of the atomic transition frequencies is sufficiently dense so that the spectral distribution can be approximated by a normalized continuous function $\rho(\w)$, where $\int \rho(\w) d\w =1$.  During the long but finite operation time of our system, the resolvable frequency can be recognized as $\sim 1/(t_f - t_o)$, as such atoms cannot be distinguished if their energy difference $d\w$ is much smaller, i.e. $d\w \ll 1/(t_f - t_o)$.  Thus the excitation of the $N\rho(\w)d\w$ atoms within the frequency strip $(\w, \w+d\omega)$ can be considered a collective one.  We define the collective atomic operator at detuning $\w$ as
\begin{equation}
    \co_{M\w} \equiv \frac{1}{\sqrt{N \rho(\omega)} d\w}\sum_{n\in (\w,\w+d\w)} \hat{\sigma}_-^{(n)} ~,
\end{equation}
where the subscript $M$ denotes memory. The collective atomic operator can be approximated as a bosonic operator as it follows approximately the bosonic commutation relation in the low excitation regime, i.e. $[\co_{M\w}, \cd_{M\w'}] \approx \frac{\delta_{\w,\w'}}{d\w} \approx \delta(\w - \w')$.  We note that the last relation is a valid definition of Dirac delta function when $d\w \rightarrow 0$.  By using the collective atomic operators, the total Hamiltonian can be rewritten as
\begin{eqnarray}\label{eq:Full_H}
H &=& \xi (\Bd_S \Bd_I + \Bo_S \Bo_I)  + \int \w \bd_{S\w}\bo_{S\w} d\w + \int \w \bd_{I\w}\bo_{I\w} d\w + \int \w \cd_{M\w} \co_{M\w} d\w \nonumber \\
&&+ \sqrt{\frac{\kappa_S}{2\pi}} \int (\Bd_S \bo_{S\w} +  \bd_{S\w}\Bo_S) d\w + \sqrt{\frac{\kappa_I}{2\pi}} \int (\Bd_I \bo_{I\w} +  \bd_{I\w}\Bo_I) d\w + G \int \sqrt{\rho(\w)} \left(\hat{B}^{\dagger}_S \co_{M\w} +  \cd_{M\w} \hat{B}_S \right) d\w~,
\end{eqnarray}
where $G \equiv \sqrt{N}g$ is the collectively enhanced atomic ensemble-cavity coupling.  We note that the continuum approximation of atomic spectral density can describe the frequency comb if each frequency strip is wider than the spacing of the frequency comb, i.e. one can find a $d\w$ such that $1/(t_f - t_o) \gg d\w \gg \Delta$.

The dynamics of the system is governed by the Langevin equations:
\begin{subequations}
\begin{eqnarray}
\dot{\Bo}_I &=& -\frac{\kappa_I}{2} \Bo_I -i\xi \Bd_S  -i \sqrt{\kappa_I} \Bin_I  \label{eq:aI_wc}\\
\dot{\Bo}_S &=& -\frac{\kappa_S}{2} \Bo_S - i \xi \Bd_I -i G \int \sqrt{\rho(\w)} \co_{M\w} d\w -i \sqrt{\kappa_S} \Bin_S \label{eq:as_wc} \\
\dot{\co}_{M\w} &=& -i \w \co_{M\w} -i G \sqrt{\rho(\w)} \Bo_S~. \label{eq:LE_spin}
\end{eqnarray}
\end{subequations}
The input and output operators are defined as $\hat{B}^{\textrm{in}}_u(t)\equiv \frac{1}{\sqrt{2\pi}}\int \bo_{u\w}(t_o) e^{-i \w (t-t_o)}d\w$ and $\hat{B}^{\textrm{out}}_u(t)\equiv \frac{1}{\sqrt{2\pi}}\int \bo_{u\w}(t_f) e^{-i \w (t-t_f)}d\w$, for $u=S, I$.

Our aim is to understand the entanglement generated between the signal and idler output fields and the atomic memory, this can be obtained from the scattering relation of the frequency modes.  First, we perform the Fourier transform of Eq.~(\ref{eq:as_wc}) and the complex conjugate of Eq.~(\ref{eq:aI_wc}),
\begin{subequations}
\begin{eqnarray}\label{eq:aI_nM}
-i \w \Bd_I [\w] &=&  -\frac{\kappa_I}{2} \Bd_I[\w] +i \xi \Bo_S[\w] + i \sqrt{\kappa_I} \Bind_I [\w] \\
-i \w \Bo_S[\w] &=& -\frac{\kappa_S}{2} \Bo_S[\w] - i \xi \Bd_I[\w] -i \sqrt{\kappa_S} \Bin_S[\w] -i G \int \int \sqrt{\rho(\w')} \co_{M\w'}(t)e^{i \w t} d\w' dt ~.  \label{eq:as_nM}
\end{eqnarray}
\end{subequations}
The input-output relation of the radiation modes remains in the standard form
\begin{subequations}\label{eq:io_Is}
\begin{eqnarray}
\Boutd_I[\w] &=& \Bind_I[\w] + i \sqrt{\kappa_I} \Bd_I[\w]  \\
\Bout_s[\w] &=& \Bin_s[\w] - i \sqrt{\kappa_S} \Bo_S[\w]~.
\end{eqnarray}
\end{subequations}
%It is obvious that all non-Markovian behavior comes from the last term.  
We can obtain the analogous input-output relation for the atomic memory by integrating Eq.~(\ref{eq:LE_spin}) from the initial time $t_o$ and to the final time $t_f$, i.e.
\begin{subequations}
\begin{eqnarray}
\co_{M\w}(t) &=& \co_{M\w}(t_o) e^{-i \w (t-t_o)} - i G \sqrt{\rho(\w)} \int_{t_o}^t \Bo_S(t') e^{-i \w (t-t')} dt'  \label{eq:spin_input} \\
&=& \co_{M\w}(t_f) e^{-i \w (t-t_f)} + i G \sqrt{\rho(\w)} \int_t^{t_f} \Bo_S(t') e^{-i \w (t-t')} dt' \label{eq:spin_output} ~.
\end{eqnarray}
\end{subequations}
Subtracting both equations, we have
\begin{equation}\label{eq:spin_io}
\co_{M\w} (t_f) e^{i \w t_f} = \co_{M\w}(t_o) e^{i \w t_o} -i G \sqrt{\rho(\w)} \int_{t_o}^{t_f} \Bo_S (t') e^{i \w t'} dt' ~.
\end{equation}
%The physical meaning of $\co_\w (t_o)$ and $\co_\w(t_f)$ is very clear: they represent the state of the collective spin mode at frequency $\w$ before and 
This relation accounts for the transformation of collective atomic modes after interacting with the cavity.  To match the standard definition of photonic input and output operators, we rescale the collective atomic operators as

\begin{equation}
\Cin_M[\w] = \sqrt{2 \pi} \co_{M\w}(t_o) e^{i \w t_o}~~~,~~~\Cout_M[\w] = \sqrt{2 \pi} \co_{M\w}(t_f) e^{i \w t_f}~.
\end{equation}
For an operation time much longer than the dynamical time scale of the system, we can set $t_o \rightarrow - \infty$ and  $t_f \rightarrow \infty$, then the integral in Eq.~(\ref{eq:spin_io}) can be recognized as the Fourier transform of $\Bo_S$.  The input-output relation of the collective atomic modes can be written as
\begin{equation}\label{eq:spin_io2}
\Cout_M[\w]=  \Cin_M[\w] -i G \sqrt{2 \pi \rho(\w)} \Bo_S [\w]~.
\end{equation}

The remaining step to obtain the input-output relation is to evaluate the last integral in Eq.~(\ref{eq:as_nM}).  By substituting in Eq.~(\ref{eq:spin_input}), we get
\begin{eqnarray}
-i \w \Bo_S[\w] &=& -\frac{\kappa_S}{2} \Bo_S[\w] - i \xi \Bd_I[\w] -i \sqrt{\kappa_S} \Bin_S[\w]  \\
&& -i G \int \int \sqrt{\rho(\w')} \co_{M\w'}(t_o) e^{-i \w'(t-t_o)} e^{i \w t} d\w' dt 
- G^2 \int \int \rho(\w') \int^t \Bo_S(t') e^{-i \w'(t-t')} dt' e^{i \w t} d\w' dt ~. \nonumber
\end{eqnarray}
The first integral can be evaluated by first integrating $t$:
\begin{eqnarray}
\int \int \sqrt{\rho(\w')} \co_{M\w'}(t_o) e^{-i \w'(t-t_o)} e^{i \w t} dt d\w'  &=&
\int  \sqrt{\rho(\w')} \co_{M\w'}(t_o) e^{i \w't_o} 2 \pi \delta(\w-\w') d\w' \nonumber \\
&=& \sqrt{2 \pi \rho(\w)} \Cin_M[\w]~.
\end{eqnarray}
For the second integral, we first rewrite the integral domain of $t'$ in terms of the Heaviside step function $\Theta(t - t')$, and then change the integral of $t$ to $\tau \equiv t-t'$:
\begin{eqnarray}
\int \int \rho(\w') \int \Theta(t-t') \Bo_S(t') e^{-i \w'(t-t')} dt' e^{i \w t} d\w' dt &=& \int \int \rho(\w') \int \Theta(\tau)  \Bo_S(t') e^{-i \w'\tau} e^{i \w \tau} d\tau e^{i \w t'} d\w' dt' \nonumber \\
&=& \int \int \rho(\w')  \Theta(\tau)   e^{-i \w'\tau} e^{i \w \tau} d\tau  d\w' \int e^{i \w t'} \Bo_s(t')dt' \nonumber \\
&=& 2 \pi \rho_+ (\w) \Bo_s [\w]~,
\end{eqnarray}
where the positive-time frequency distribution $\rho_+(\w)$ is defined as
\begin{equation}\label{eq:rho_plus}
\rho_+(\w) \equiv \frac{1}{2 \pi} \int \int \rho(\w')  \Theta(\tau)   e^{-i \w'\tau} e^{i \w \tau} d\tau  d\w' 
=\frac{1}{2 \pi} \int_0^\infty \int \rho(\w') e^{-i (\w'-\w)\tau} d\w' d\tau~.
\end{equation}
Putting all these into Eq.~(\ref{eq:as_nM}), we have
\begin{equation}\label{eq:as_nM2}
-i \w \Bo_S[\w] = \left(-\frac{\kappa_S}{2} - G^2 2 \pi \rho_+(\w)\right) \Bo_s[\w] - i \xi \Bd_I[\w] -i \sqrt{\kappa_S} \Bin_S[\w] - i G \sqrt{2 \pi \rho(\w)} \Cin_M[\w]~.
\end{equation}

By combining Eqs.~(\ref{eq:aI_nM}), (\ref{eq:as_nM2}), (\ref{eq:io_Is}), and (\ref{eq:spin_io2}), the input-output relation can be obtained as (i.e. Eq.~(2) in the main text)
\begin{subequations}
\begin{eqnarray}
\Boutd_I[\w] &=& T_{I I}(\w) \Bind_I[\w] + T_{IS}(\w) \Bin_S[\w] + T_{IM}(\w) \Cin_M[\w] \\
\Bout_S[\w] &=& T_{S I}(\w) \Bind_I[\w] + T_{SS}(\w) \Bin_S[\w] + T_{SM}(\w) \Cin_M[\w] \\
\Cout_M[\w] &=& T_{M I}(\w) \Bind_I[\w] + T_{MS}(\w) \Bin_S[\w] + T_{MM}(\w) \Cin_M[\w] ~,
\end{eqnarray}
\end{subequations}
or in the matrix form $\bm{\hat{B}}^\textrm{out}= \bm{T}\bm{\hat{B}}^\textrm{in}$, where $\bm{\hat{B}}^\textrm{in/out}\equiv (\hat{B}^{\textrm{in/out} \dag}_I[\w]~\hat{B}^{\textrm{in/out}}_S[\w]~\hat{C}^{\textrm{in/out}}_M[\w] )^\textrm{T}$. 
The matrix $\bm{T}$, which contains the scattering amplitudes, can be calculated by
\begin{equation}
\bm{T}(\w) \equiv \bm{I}_3 + i \tilde{\bK}^\textrm{T}(\w) \left(\w \bm{I}_2- \tilde{\bH}(\w) \right)^{-1} \tilde{\bK}(\w)~,
\end{equation}
where $\bm{I}_k$ is the $k \times k$ identity matrix;
\begin{equation}
\tilde{\bH}(\w) \equiv \begin{pmatrix}  - i \kappa_I/2 & -\xi \\
\xi &  -i \kappa_S/2 -i G^2 2 \pi \rho_+(\w) 
\end{pmatrix}~~;~~
\tilde{\bK}(\w) \equiv \begin{pmatrix}
i \sqrt{\kappa_I}  &  0 & 0 \\
0 & -i \sqrt{\kappa_S} & -i G \sqrt{2\pi \rho(\w)}
\end{pmatrix}~.
\end{equation}

\subsection{System stability}

The formalism outlined in the last section applies to any spectral distribution of the atomic ensemble.  For simplicity, in this stability analysis we consider a Lorentzian spectrum, $\rho(\w) = \frac{1}{2\pi} \frac{\Gamma}{\w^2 + (\Gamma/2)^2}$, which is characterized by only the inhomogeneous broadening $\Gamma$. The merit of Lorentzian spectrum is that simple analytical results can be obtained for various properties of interest (e.g. Eq.~(3) in the main text), which are generally challenging to obtain for a non-flat spectrum of atoms. This can provide us valuable physical insights regarding the properties of the system. 

To be more specific, we recall that the dynamics of a quantum system coupling to a non-Markovian bath is equivalent to one that interacts with dissipative quasi-modes that couple to Markovian baths (see Refs.~\cite{2001PhRvA..64e3813D, 2018PhRvL.120c0402T}).  The simplicity of a Lorentzian spectrum is that only one quasi-mode is needed to account for the non-Markovian dynamics.  For our setup, we consider the following equivalent system, which instead of a Lorentzian atomic ensemble the signal mode is coupled to a resonant but damping quasi-mode, $\ao_Q$,
\begin{eqnarray}\label{eq:Equiv_H}
H &=& \xi (\Bd_S \Bd_I + \Bo_S \Bo_I) + G(\Bd_S \ao_Q + \ad_Q \Bo_S) + \int \w \bd_{S\w}\bo_{S\w} d\w + \int \w \bd_{I\w}\bo_{I\w} d\w + \int \w \bd_{Q\w} \bo_{Q\w} d\w \nonumber \\
&&+ \sqrt{\frac{\kappa_S}{2\pi}} \int (\Bd_S \bo_{S\w} +  \bd_{S\w}\Bo_S) d\w + \sqrt{\frac{\kappa_I}{2\pi}} \int (\Bd_I \bo_{I\w} +  \bd_{I\w}\Bo_I) d\w + \sqrt{\frac{\Gamma}{2 \pi}} \int \left(\ad_Q \bo_{Q\w} +  \bd_{Q\w} \ao_Q \right) d\w~,
\end{eqnarray}
where signal mode and the quasi-mode are interacting through a beam-splitter interaction with strength $G$, and damping of quasi-mode is modelled as a coupling to a Markovian bath that consists of a continuum of bosonic modes at different detuning, $\bo_{Qw}$.  As we will see, the quasi-mode damping rate $\Gamma$ coincides with the inhomogeneous broadening in our setup.

The equation of motion of the cavity and quasi- modes are given by
\begin{subequations}\label{eq:Equiv_eom}
\begin{eqnarray}
\dot{\Bo}_I &=& -\frac{\kappa_I}{2} \Bo_I -i\xi \Bd_S  -i \sqrt{\kappa_I} \Bin_I \label{eq:Equiv_eom1} \\
\dot{\Bo}_S &=& -\frac{\kappa_S}{2} \Bo_S - i \xi \Bd_I -i G \ao_Q -i \sqrt{\kappa_S} \Bin_S  \label{eq:Equiv_eom2}\\
\dot{\ao}_Q &=& -\frac{\Gamma}{2}\ao_Q - i G \Bo_S -i \sqrt{\Gamma} \Bin_Q~, \label{eq:Equiv_eom3}
\end{eqnarray}
\end{subequations}
where $\hat{B}^{\textrm{in}}_Q(t)\equiv \frac{1}{\sqrt{2\pi}}\int \bo_{Q\w}(t_o) e^{-i \w (t-t_o)}d\w$.  By integrating Eq.~(\ref{eq:Equiv_eom3}) and substituting into Eq.~(\ref{eq:Equiv_eom2}), we get
\begin{eqnarray}
    \dot{\Bo}_S &=& -\frac{\kappa_S}{2} \Bo_S - i \xi \Bd_I - G^2 \int^t e^{-\frac{\Gamma}{2} (t-t')} \Bo_S(t') dt' -G\sqrt{\Gamma}\int^t e^{-\frac{\Gamma}{2} (t-t')} \Bin_Q(t') dt' -i \sqrt{\kappa_S} \Bin_S \\
    &=& -\frac{\kappa_S}{2} \Bo_S - i \xi \Bd_I - G^2 \int^t \int \frac{\Gamma}{2\pi}\frac{e^{-i\w (t-t')}}{\w^2 + (\Gamma/2)^2} d\w \Bo_S(t') dt' -G\int \sqrt{\frac{\Gamma}{2\pi}}\frac{1}{-i\w +\Gamma/2} \bo_{Q\w}(t_o)e^{-i\w (t-t_o)}  d\w -i \sqrt{\kappa_S} \Bin_S~. \nonumber \\ \label{eq:BS_equiv}
\end{eqnarray}

On the other hand, in our setup we can derive the equation of motion of signal mode by integrating Eq.~(\ref{eq:LE_spin}) and substituting into Eq.~(\ref{eq:as_wc}):
\begin{equation}
    \dot{\Bo}_S = -\frac{\kappa_S}{2} \Bo_S - i \xi \Bd_I - G^2 \int^t \int \rho(\w) e^{-i\w(t-t')} \Bo_S(t') d\w dt' -i G \int \sqrt{\rho(\w)} \co_{M\w}(t_o)e^{-i \w (t-t_o)} d\w  -i \sqrt{\kappa_S} \Bin_S~. \label{eq:BS_origin}
\end{equation}
For our choice of Lorentzian distribution $\rho(\w) = \frac{1}{2\pi} \frac{\Gamma}{\w^2 + (\Gamma/2)^2}$, it is obvious that Eqs.~(\ref{eq:BS_origin}) and (\ref{eq:BS_equiv}) are equivalent, up to a frequency dependent phase in each bath mode which does not affect the dynamics of $\Bo_S$ (i.e. $\sqrt{\frac{\Gamma}{2\pi}}\frac{1}{-i\w +\Gamma/2}=i\sqrt{\rho(\w)}e^{i\phi}$ for some phase factor $\phi$.)  Since Eqs.~(\ref{eq:aI_wc}) and (\ref{eq:Equiv_eom1}) are identical, we can conclude that the cavity mode dynamics is the same in our setup as the quasi-mode model.

As a result, the stability of our setup can be analyzed by studying the equivalent equation of motion (\ref{eq:Equiv_eom}).  We first rewrite Eq.~(\ref{eq:Equiv_eom}) in the matrix form
\begin{equation}
    \dot{\bm{B}}=\bm{M}\bm{B}+\bm{B^\textrm{in}}~~,~~\textrm{where } 
    \bm{M}\equiv \begin{pmatrix}-\kappa/2 & i \xi & 0 \\-i \xi & -\kappa/2 & -iG \\ 0 & -iG & -\Gamma/2 \end{pmatrix}~,
\end{equation}
$\bm{B}\equiv (\Bd_I~\Bo_S~\ao_Q)^\textrm{T}$ and $\bm{B}^\textrm{in}\equiv (i \sqrt{\kappa}\Bind_I~-i\sqrt{\kappa}\Bin_S~-i\sqrt{\Gamma}\Bin_Q)^\textrm{T}$.  We have assumed $\kappa_S = \kappa_I \equiv \kappa$ for simplicity.  The system is stable if and only if all eigenvalues of $\bm{M}$ have strictly negative real parts.  This can be verified by checking the Routh-Hurwitz stability criterion of the eigenvalue equation of $\bm{M}$ \cite{2011Csf}.  We find that the system is stable when the parametric drive strength $\xi$ obeys
\begin{equation}
    \frac{4 \xi^2}{\kappa^2} < \min\left\{ \sC+1, \Big(\frac{\sC}{2}x + x+1 \Big) (x+1)\right\}~,
\end{equation}
where $\sC \equiv 4 G^2/\kappa \Gamma$ is the atomic ensemble-cavity cooperativity and $x\equiv \Gamma/\kappa$ is the ratio of inhomogeneous broadening to cavity damping rate.

\subsection{Entanglement rate}
We quantify the performance of our system by the rate of entanglement generated between the atomic ensemble memory and the output radiation, i.e. Eq.~(4) in the main text. Here we outline the procedure of computing this quantity. We first recall that, according to the scattering relation Eq.~(2) in the main text, the idler and signal frequency modes $\Bout_I[-\w]$ and signal $\Bout_S[\w]$ and the collective atomic mode $\Cout_M[\w]$ form a three-mode entangled state that is equivalent to a TMS state passing through a beam splitter.  The scattering relation also tells us that frequency modes with different $\w$ do not couple to other others.  As a result, the total state of the output radiation and atomic memory is the tensor product of the state of each frequency mode.  

Secondly, because the entangled state of each frequency mode is Gaussian, the entanglement can be quantified by established continuous-variable entanglement measures.  We employ the computable measure proposed in Ref.~\cite{2017PhRvA..96f2338T}, which is generally the lower bound of entanglement of formation.  Moreover, as we will see that the state of each frequency mode is balanced, this entanglement measure is exactly the entanglement of formation in our setup.  We note that the entanglement rate calculated by our method is the maximum amount of entanglement generated per unit time by the system; this rate will not be increased by any manipulation of the output, e.g. post-processing or non-Gaussian operations.

We computed the entanglement rate for two cases: (i) between the atomic ensemble and all radiation outputs, i.e. both idler and signal, and (ii) between the spin ensemble and idler output only, i.e. signal output is discarded.  In case (i), because the three-mode state is pure, the entanglement of formation coincides with the von Neumann entropy of the atoms:
\begin{equation}\label{eq:vE}
\sE(\w) = S(\rho_c(\w)) = \big(\bar{n}(\w)+1\big) \log_2 \big(\bar{n}(\w)+1\big) - \bar{n}(\w) \log_2 \bar{n}(\w)~.
\end{equation}
$\bar{n}(\w)$ is the mean excitation number of collective atomic mode at frequency $\w$; this can be obtained from the scattering relation (Eq.~(2) in the main text):
\begin{equation}
\bar{n}(\w) = |T_{M I}(\w)|^2 = \sin^2 \theta_2(\w) \sinh^2 r(\w)~.
\end{equation}
The last expression relates to the effective TMS strength $r(\w)$ and relevant BS angle $\theta_2(\w)$ in the equivalent circuit model in Fig.~1(c).

In case (ii), we first construct the covariance matrix \cite{weedbrook_gaussian_2012} of the two-mode state between idler mode $\Bout_I [-\w]$ and collective atomic mode $\Cout_M[\w]$. This can be obtained by using the input-output relation in main text Eq.(2), and the fact that the input photonic and initial atomic states are respectively vacuum and ground state, i.e. $\Bin_S |\textrm{vac}\rangle = \Bin_I |\textrm{vac}\rangle = \Cin_M |\textrm{gnd}\rangle =0$. By expressing the covariance matrix in the standard form of two-mode Gaussian state \cite{Duan:2000ts}, we get
\begin{equation}
V = \begin{pmatrix}
\frac{a}{2} & 0 & \frac{b}{2} & 0 \\
0 & \frac{a}{2} & 0 & -\frac{b}{2} \\
\frac{b}{2} & 0 & \frac{c}{2} & 0 \\
0 & -\frac{b}{2}& 0 & \frac{c}{2} \\
\end{pmatrix}
\end{equation}
where the entries are given by
\begin{subequations}
\begin{eqnarray}
a & \equiv & 2 \left(|T_{IS}(\w)|^2 + |T_{IM}(\w)|^2 \right) +1 = 2 \sinh^2 r(\w) +1 \\
b & \equiv & 2 |T_{MI}(\w) T^\ast_{II}(\w)| = 2\sin\theta(\w) \cosh r(\w) \sinh r(\w)\\
c & \equiv & 2 |T_{MI}(\w)|^2 +1 = 2 \sin^2 \theta_2 (\w) \sinh^2 r(\w)+1~.
\end{eqnarray}
\end{subequations}
This two-mode state is balanced in the sense that the anti-diagonal blocks are proportional to a Pauli $Z$ matrix. Ref.~\cite{2017PhRvA..96f2338T} shows that the entanglement of formation of such state is exactly given by
\begin{equation}\label{eq:EF}
\sE(\w) = \sE_F(\w) = \big( \sinh^2 r_0(\w)+1\big) \log_2 \big(\sinh^2 r_0(\w)+1\big) - \sinh^2 r_0(\w) \log_2 \sinh^2 r_0(\w)~,
\end{equation}
where the effective squeezing parameter $r_0$ can be obtained as
\begin{equation}
e^{2 r_0 (\w)} = \frac{1+ \sin\theta_2(\w) \tanh r(\w)}{1- \sin\theta_2(\w) \tanh r(\w)}~.
\end{equation}

As a reality check, we verify that the entanglement in case (i) is always higher than that in case (ii), because discarding a part must not increase entanglement.  We first note that both Eqs.~(\ref{eq:vE}) and (\ref{eq:EF}) have the form of entropy, so their magnitude can be compared using the argument $\bar{n}(\w)$ and $\sinh^2 r_0(\w)$.  Then it can be easy to see that
\begin{equation}\label{eq:SgeqEF}
\bar{n}(\w) = \sin^2 \theta_2(\w) \sinh^2 r(\w) = \frac{\sin^2 \theta_2(\w) \sinh^2 r(\w)}{\cosh^2 r(\w) - \sinh^2 r(\w)} 
 \geq  \frac{\sin^2 \theta_2(\w) \sinh^2 r(\w)}{\cosh^2 r(\w) - \sin^2 \theta_2(\w) \sinh^2 r(\w)} = \sinh^2 r_0 (\w)~,
\end{equation}
and so $S(\rho_c(\w)) \geq \sE_F(\w)$, as expected.

\subsection{Photon retrieval efficiency}

The entanglement generated between idler output and atomic memory can be retrieved by different means, for instance, by coupling the atomic ensemble to superconducting resonators and transducing to microwave photons. In this section, we discuss the efficiency of a particular way of retrieval that the stored quantum state is converted back to the optical signal photon output.

In the retrieval stage, the parametric drive will be switched off to avoid unwanted mixing with the atomic state that stores the entanglement.  We consider a general situation where the inhomogeneously broadened quantum memory will be re-focused after certain storage time, e.g. due to implementation of atomic frequency combs.  The stored quantum state will then be converted to photons in the signal cavity, and subsequently will leak to the signal waveguide.  Such a process can be described by the scattering relation Eq.~(2) in the main text with $\xi =0$.  We note that in employing these scattering relations, the initial state of the atoms (characterized by $\Cin_M[\w]$) is no longer the ground state but rather the stored entangled state. To focus our investigation on the retrieval process, we assume memory does not decohere during the storage, so the $\Cin_M[\w]$ in the retrieval stage can be chosen as the $\Cout_M[\w]$ of the entanglement generation stage.  

Since the idler is decoupled when $\xi =0$, the steady-state scattering relation in the retrieval process involves only a mixing of signal waveguide mode $\Bin_S[\w]$ and collective atomic mode $\Cin_M[\w]$, both at the detuning $\w$. The process is thus analogous to a beam splitter with a detuning dependent angle.  The efficiency of retrieving the state of the collective atomic mode is thus determined by the scattering amplitude $T_{SM}$, i.e.
\begin{equation}
    |T_{SM}(\w)|^2 = \frac{G^2 \Gamma \kappa}{(G^2+\kappa \Gamma/4-\w^2)^2 + (\kappa/2 + \Gamma/2)^2\w^2}~.
\end{equation}
From this explicit expression, a perfect retrieval (i.e. $|T_{SM}|^2=1$) requires two conditions: first, the frequency modes have zero detuning $\w=0$, and second, the system parameters satisfy $G^2 = \kappa \Gamma/4$, which is the impedance matching condition.  For our broadband quantum memory, perfect retrieval is impossible for all frequency modes with any set of system parameters.

\begin{figure}[ht]
  \centering
    \includegraphics[width=.5\linewidth]{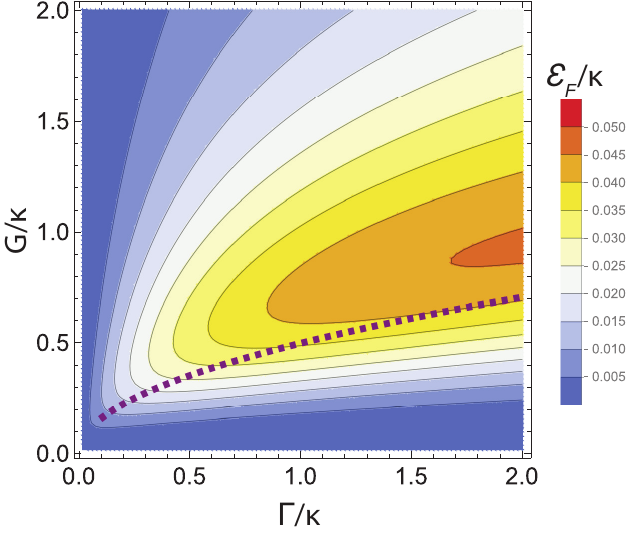}
\caption{Normalized entanglement rate between the idler output at the entanglement generation stage and the signal output at the retrieval stage.  The system parameters are identical to Fig.~2(a) in the main text.  Dotted line represents the impedance matching condition. 
} 
\label{fig:retrieval}
\end{figure}

Instead of looking for designs and controls to achieve broadband perfect retrieval, which are expected to be challenging to implement and outside the scope of the present work, we study how the imperfect retrieval affects the retrieved entanglement.  In analogy to Fig.~2(a) in the main text, we show in Fig.~\ref{fig:retrieval} the entanglement rate between idler output and the retrieved signal output.  This is computed with the overall scattering relation, which is obtained by cascading the scattering relation in Eq.~(2) for the retrieval stage after that for the entanglement generation stage.

\begin{figure}[ht]
  \centering
  \begin{subfigure}{.5\textwidth}
  \centering
  \includegraphics[width=.9\linewidth]{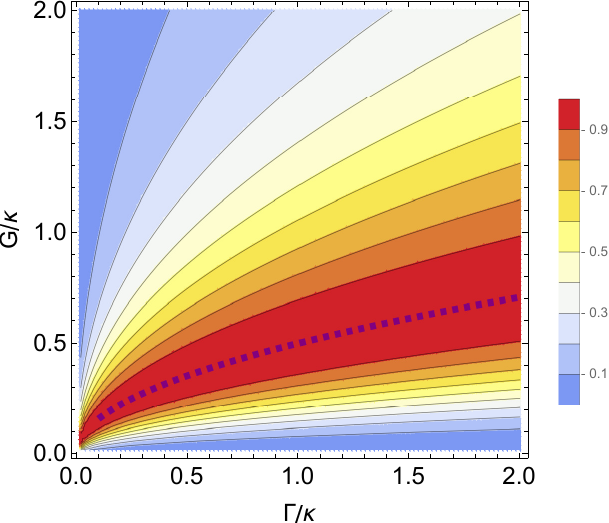}
  \end{subfigure}%
    \begin{subfigure}{.5\textwidth}
    \centering
  \includegraphics[width=.9\linewidth]{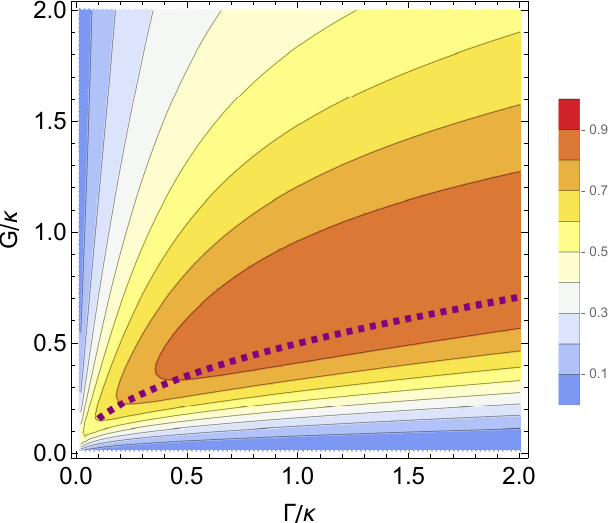}
  \end{subfigure}
\caption{(Left) Retrieval efficiency for the resonant mode, i.e. $|T_{SM}(0)|^2$, for the system parameters in the main text Fig.~2(a). The efficiency is maximum around the vicinity of the impedance matching condition (dotted line). (Right) Ratio of retrieved entanglement rate to the generated entanglement rate for the same set of system parameters. High retrieval ratio is achieved for a wide range of system parameters, not limited to the vicinity of impedance matching.
} 
\label{fig:retrieval_efficiency}
\end{figure}

With the same generic, weakly driven (i.e. $\xi = 0.1 \kappa$) system parameters as in the main text Fig.~2(a), Fig.~\ref{fig:retrieval} shows that the impedance matching condition is not optimal for retrieving entanglement from our broadband quantum memory. This is the consequence of two effects: first, as shown in the main text Fig.~2(a), the entanglement generation rate is not optimal at impedance matching; second, entanglement is generated across a spectrum of frequency modes, so the parameters that optimize the resonant mode retrieval do not generally optimize the collective retrieval efficiency of all frequency modes. On the other hand, the retrieved entanglement rate remains considerable for a wide range of system parameters when comparing to the entanglement generated in the atomic memory.  As shown in Fig.~\ref{fig:retrieval_efficiency}, over 80\% of entanglement can be retrieved in this generic example.  This is because of the broadband nature of the system (i.e. characteristic frequency range is determined by the cavity-waveguide coupling $\kappa$ and inhomogeneous broadening $\Gamma$), and also the robustness of CV entanglement against loss \cite{physreva.83.024301}.  

\begin{figure}[ht]
  \centering
  \begin{subfigure}{.5\textwidth}
  \centering
  \includegraphics[width=.9\linewidth]{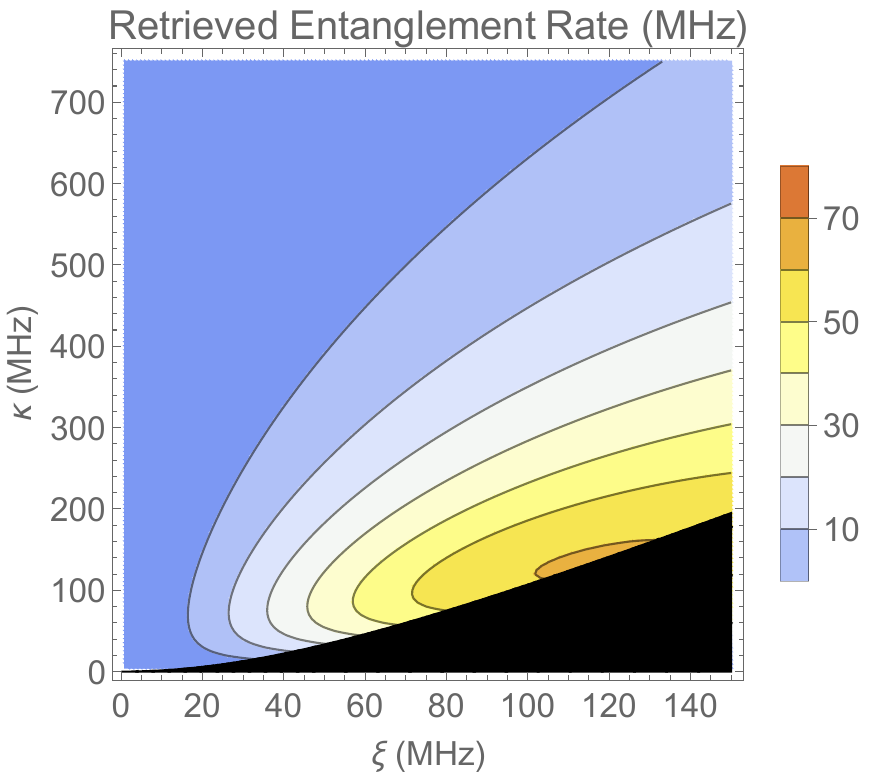}
  \end{subfigure}%
    \begin{subfigure}{.5\textwidth}
    \centering
  \includegraphics[width=.9\linewidth]{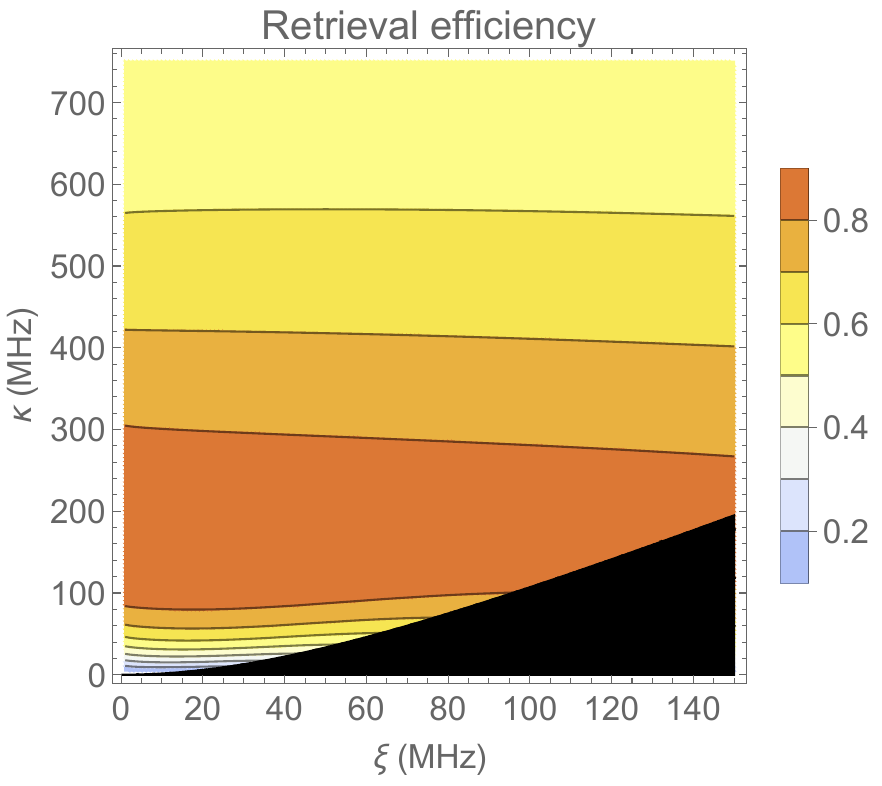}
  \end{subfigure}
\caption{(Left) Retrieved entanglement rate for a system with parameters given in the main text Table I. (Right) Retrieval efficiency of the same system.
} 
\label{fig:retrieval_Fig4}
\end{figure}

We note that the high retrieval rate is not only achievable in the weak drive regime (i.e. $\xi \ll \Gamma, \kappa$).  In Fig.~\ref{fig:retrieval_Fig4}, we also show the retrieved entanglement rate for a system with realistic parameters given in the main text Table I. By varying the cavity decay rate $\kappa$ and parametric drive strength $\xi$, our simulation shows that the retrieved entanglement rate can be higher than 60 MHz. The retrieval efficiency, which is defined as the ratio of the retrieved entanglement rate to the entanglement generation rate (i.e. Fig.~4(a) in the main text), is over 80\% for a wide range of system parameters, covering both the weak and strong driving regimes. This further supports that our in-situ quantum memory scheme allows high retrieval efficiency without precisely tuning to the impedance matching condition.

\subsection{Tripartite entanglement}

Our system will generate a tri-partite entangled state between the respective frequency modes of idler output, signal output, and memory.  The state is equivalent to the resultant state after a mode of a two-mode squeezed state interact with another mode in the ground state through a beam splitter.  To verify this entangled state is truly tri-partite, we first note that the phases of the two-mode squeezing and beam splitter does not affect the entanglement, therefore we examine the case where both phases are zero for simplicity. The state will then be completely characterized by the two-mode squeezing strength $r$ and the beam-splitting angle $\theta$.  The covariance matrix of a general state in this class will take the form
\begin{equation}
\begin{pmatrix}
    a & 0 & b & 0 & c & 0 \\
    0 & a & 0 & -b & 0 & -c \\
    b & 0 & d & 0 & e & 0 \\
    0 & -b & 0 & d & 0 & -e \\
    c & 0 & e & 0 & f & 0 \\
    0 & -c & 0 & -e & 0 & f 
\end{pmatrix}~,
\end{equation}
where $a\equiv \frac{1}{2}\cosh 2r$, $b\equiv \cos\theta \cosh r \sinh r$, $c\equiv -\sin\theta \cosh r \sinh r$, $d\equiv \frac{1}{2}(\cos^2\theta \cosh 2r +\sin^2 \theta)$, $e \equiv - \sin\theta \cos\theta \sinh^2 r$, $f \equiv d\equiv \frac{1}{2}(\sin^2\theta \cosh 2r +\cos^2 \theta)$.  A pure state is genuinely multi-partite entangled if every partition of the parties are entangled.  In our case, there can be three different partitions: \{signal, idler + memory\}, \{idler, signal + memory\}, and \{memory, idler + signal\}.  The entanglement of all three partitions can be verified by showing the reduced state of each mode to be mixed.  It is easy to see that $a>1/2$, $d>1/2$, and $f>1/2$ for any non-trivial squeezing parameter $r$ and beam-splitting angle $\theta$.  This implies the reduced state of all three modes are thermal state with non-zero excitation \cite{weedbrook_gaussian_2012}, and thus mixed. 

\subsection{Cavity design and experimental parameters}

\begin{figure}[ht]
  \centering
  \includegraphics[width=\linewidth]{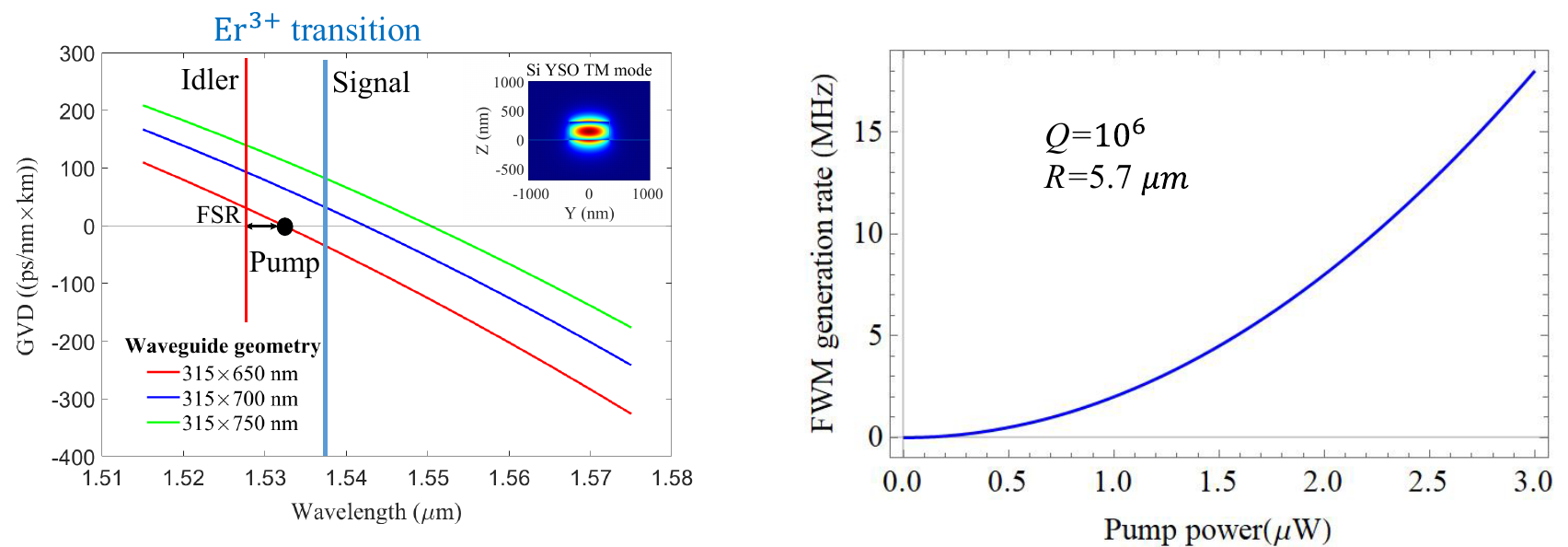} 
\caption{Experimental resonator design. (Left) Zero-dispersion wavelength for four-wave-mixing phase matching by changing the waveguide geometry. The designed waveguide thickness is 315 and the width is 650 nm. (Right) Spontaneous four-wave-mixing (sFWM) generation rate as a function of pump power in a Si ring resonator with radius of 5.7$\mu$m and Q of $10^6$.
} 
\label{fig:cavity design}
\end{figure}

Here we show detailed calculations of the experimental configuration of a micro-ring resonator on a $\rm {}^{167} Er^{3+}$:$\rm Y_2SiO_5$ crystal based on the spectroscopic properties of this quantum memory material \cite{rancic_coherence_2018}. The $\rm {}^{167} Er^{3+}$ ions are evanescently coupled to the optical field in the silicon waveguide as shown in Fig.~3 and Fig.~\ref{fig:cavity design}(Left) inset. The inset shows the mode profile of silicon waveguide on top of a Er$^{3+}$:YSO substrate. From the waveguide transerve-magnetic (TM) mode simulation, we engineer the zero dispersion point at pump frequency so that the signal and idler photons will be phase matched at one free-spectral range (FSR) away from the pump frequency, as shown in Fig.~\ref{fig:cavity design}(Left). In this nonlinear cavity, entangled signal and idler photon pairs can be generated in a spontaneous four-wave-mixing (sFWM) process with a low optical pumping power. According to \cite{azzini_ultra-low_2012}, we estimate the sFWM generation rate as a function of pump power for our system in Fig.~\ref{fig:cavity design}(Right),
\begin{equation}
    \xi_{FWM}=(\gamma P_p 2\pi R)^2 (\frac{Q v_g}{\nu_p \pi R})^3 \frac{v_g}{4\pi R} P^2_{pump}=2 ~\rm (\frac{MHz}{\mu W^2}) P^2_{pump} 
\end{equation}

\noindent where $\gamma=\frac{\omega_0 n_2}{c A_{\rm eff}}$. $A_{\rm eff}$ is the effective mode area, $P_p$ is pump power, $v_g$ is group velocity and $R$=5.7 $\mu$m is the design ring radius. We operate at a low pump power ($<$10$\mu$W) for cryogenic experiment. 

The generated signal photon is coupled to the 1539 nm optical transition of $\rm {}^{167} Er^{3+}$ ensemble, transferring the photon-photon entanglement to photon-memory entanglement. The left insect of Fig.~3 and Fig. S5 shows $\rm {}^{167} Er^{3+}$ hyperfine levels, their spacings and relative detunings between relevant optical transitions near 1539 nm. The ground state and excited state are spitted into 16 hyperfine states denoted by nuclear spin quantum number $|m_s\rangle$. With 7 T magnetic field applied along D1 axis of $\rm {}^{167} Er^{3+}:YSO$ crystal, $\rm\Delta m_s=0, \pm 1$ transitions  between the ground and excited hyperfine states become spectrally resolved from each other. In order to initialize the spin states with optical pumping and isolate a two level system for AFC quantum memory, both $\Delta m_s=0$ (like-like) transitions and $\Delta m_s=\pm 1$ cross transitions should be spectrally separated by more than the cavity linewidth. The splitting between ground state $m_s=7/2$ and $m_s=5/2$ hyperfine levels at 7 T magnetic field is 796 MHz \cite{rancic_coherence_2018}. A typical Si ring resonator with a quality factor $Q=10^6 $ gives a cavity linewidth that is comparable to inhomogeneous broadening of the Er dopants $\Gamma_{\rm {inh}}=150$ MHz, and is much smaller than the $\rm {}^{167} Er^{3+}$ ground state hyperfine splitting. This configuration allows us to couple signal photons dominantly to the $m_s=7/2$ like-like transition (there is a partial overlap with the $m_s=5/2$ like-like transition, which we discuss below) while efficient optical pumping via the $\Delta m_s = +1$ cross transition is still achievable with a more intense pump.
\begin{figure}[ht]
  \centering
  \includegraphics[width=\linewidth]{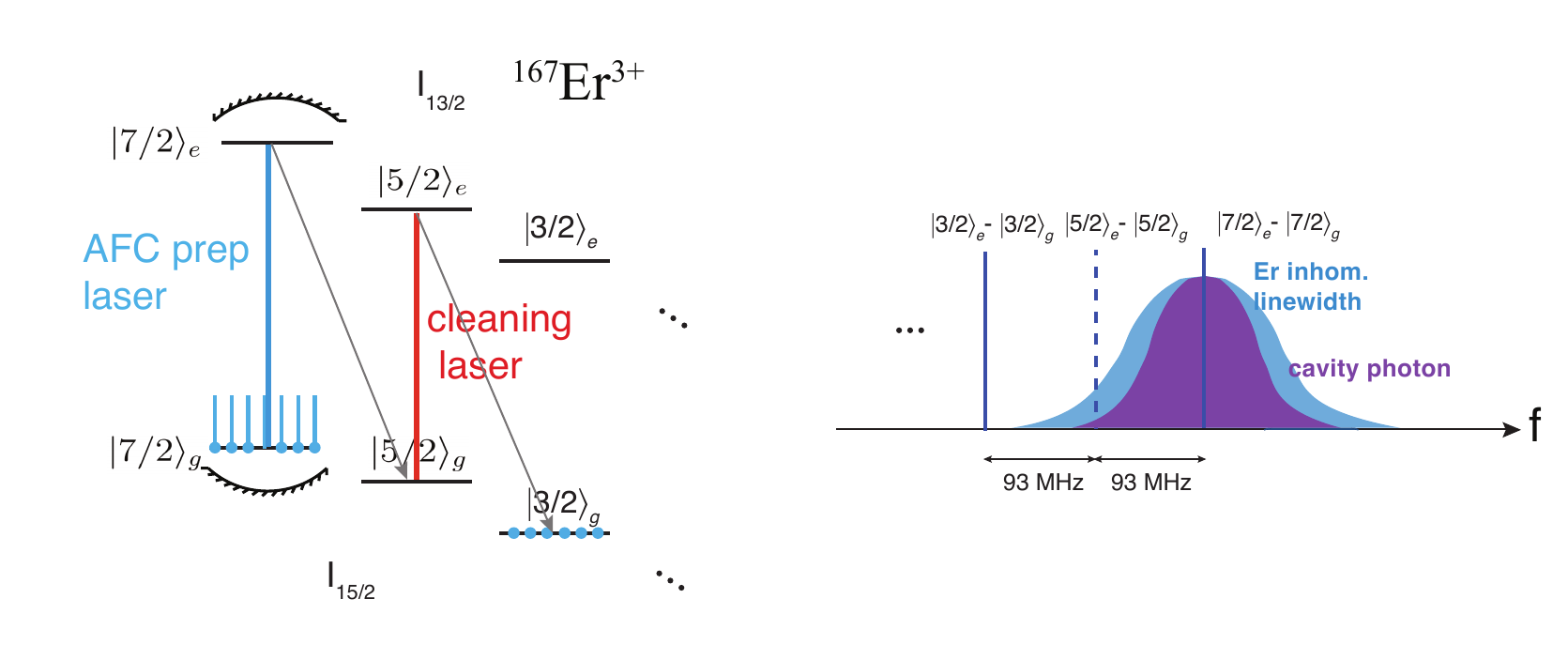} 
\caption{Hyperfine levels and optical transitions of $\rm {}^{167} Er^{3+}:YSO$ with respect to the cavity photons. Left panel illustrates the proposed atomic frequency comb preparation procedure, which uses a cleaning laser resonant with the $|5/2\rangle_e-|5/2\rangle_g$ transitions to deplete residual population in the $|5/2\rangle_g$ ground level.
} 
\label{fig:cavity design.}
\end{figure}

Ideally, we would like the signal photons predominantly coupled to the $m_s=7/2$ like-like transition, and not to other adjacent optical transitions in $\rm {}^{167} Er^{3+}$ for future spin-wave memory realization. Figure S5 shows in details the energy levels of $\rm {}^{167} Er^{3+}:YSO$ and their spacings in the optical spectrum. The immediately adjacent optical transition is the $|5/2\rangle_e-|5/2\rangle_g$ transition, which is red-detuned by 93 MHz from the $|7/2\rangle_e-|7/2\rangle_g$ AFC transition \cite{rancic_coherence_2018}. The next adjacent transition, $|3/2\rangle_e-|3/2\rangle_g$, is 186 MHz red-detuned, which would have very small coupling to the signal photons given the coupling strength G = 100 MHz. To suppress potential coupling to the $|5/2\rangle_e-|5/2\rangle_g$ transition during photon storage, we can implement a cleaning laser around resonance of the $|5/2\rangle_e-|5/2\rangle_g$ transition while the AFC is created on the $|7/2\rangle_e-|7/2\rangle_g$ transition. This procedure is illustrated in the left panel of Fig.~S5, which would deplete most of the population in the $|5/2\rangle_g$ level, thus suppresses coupling to the cavity photons. The residual atomic population will end up in the $|3/2\rangle_g$ or other hyperfine ground states, which do not contribute to the photon storage. We should also note that for the optical two-level AFC storage, this requirement of only coupling to one hyperfine spin level is relaxed; any residual population in the $|5/2\rangle_g$ level and coupling to it can be considered as an expansion of the AFC bandwidth at the lower frequency tail of the Er transition, which would not degrade the optical AFC memory performance.

The simulated mode volume of our designed ring resonator is 
$\rm V_{YSO}= \frac{\int_{V_{YSO}} \epsilon_{YSO} |E(r)|^2}{\epsilon_{max} |E_{max}|^2}=0.7 \rm \mu m^3$, 
which leads to $1\times10^5$ ions in cavity with a 17 ppm doping concentration. With a single-ion cavity coupling $$g_0=\mu \sqrt{\frac{\omega}{2\hbar V_{mode}\epsilon_{max}}}$$, and a transition dipole moment $\mu=2.07\times 10^{-32} \rm C m$ \cite{mcauslan_strong-coupling_2009}, the collective coupling of the entire ensemble is
\begin{equation}
    (\int g^2_{ion}(\vec{r}) \rho dV)^{1/2}=(\frac{\rho \omega \mu^2}{2\hbar \epsilon_{YSO}} \frac{\int_{V_{YSO}} \epsilon_{YSO} |\hat{\mu}\cdot\vec{E}(r)|^2}{\int_{V} \epsilon(r) |\vec{E}(r)|^2})^{1/2}=2 \pi \times 0.173\rm GHz.
\end{equation}

%The fraction of energy of TM mode in substrate is $\beta=\frac{\int_{V_{YSO}} \epsilon_{YSO} |\hat{\mu}\cdot\vec{E}(r)|^2}{\int_{V} \epsilon(r) |\vec{E}(r)|^2})^{1/2}=0.086$.%
The collective cooperativity is calculated as 
\begin{equation}
\sC=\frac{|W(0)|}{\kappa/2}
\end{equation}
where $W(\omega)=G^2 \int \frac{\rho(\omega') d\omega'}{\omega-\omega'}$ is the atomic absorption for Er ion inhomogeneous distribution $\rho(\omega')$. When an AFC is prepared in the ensemble, the magnitude of $G$ is reduced by square root of the AFC finesse, and the cooperativity is reduced by the AFC finesse. For a finesse of 3 in the proposed experiment, the effective cooperativity for a $\kappa$=$2\pi\times$100 MHz is 6.

\end{document}